\documentclass{iau} 
\usepackage{graphicx}
\usepackage{hyperref}
\usepackage{bm}

\title[S347.~~Early science with MANIFEST] %% give here short title %%
{Key early science with MANIFEST on GMT}

\author[Matthew Colless]   %% give here short author list %%
{Matthew Colless$^1$}

\affiliation{$^1$Research School of Astronomy and Astrophysics,
                         Australian National University \\
                         Canberra, ACT 2611, Australia \\
                         email: {\tt matthew.colless@anu.edu.au}}

\pubyear{2018}
\volume{347}  %% insert here IAU Symposium No.
\jname{Early Science with ELTs}
\editors{N.\ Przybilla, K.\ Pollard \& A.\ Calamida, eds.}

\begin{document}

\maketitle

\begin{abstract}
The MANIFEST fibre system provides a highly versatile feed for the GMACS
and G-CLEF first-light spectrographs on the Giant Magellan Telescope
(GMT). Combining these low- and high-resolution optical spectrographs
with the wide field of view (up to 20~arcmin), high multiplex, and
integral field capabilities provided by MANIFEST enables science
programs that are not achievable with other extremely large
telescopes. For galactic archaeology and near-field cosmology studies of
Local Group galaxies, MANIFEST and G-CLEF can obtain up to 40
simultaneous high-resolution optical spectra over a wide field, and so
produce detailed kinematic and chemical maps of the stellar populations
out to large radius in galaxies covering a broad range of masses and
morphologies. For galaxy evolution studies, MANIFEST and GMACS can
combine a survey of galaxies at the epoch of peak star formation with a
study of the flows of gas between galaxies and the circumgalactic
medium, mapping both the emission from hot gas using integral field
spectroscopy and the absorption from cold gas with multi-object
spectroscopy of background sources. These programs will feature strongly
in the early science goals for GMT. 
\keywords{telescopes, instrumentation: spectrographs, techniques: spectroscopic}
\end{abstract}

\noindent The Giant Magellan Telescope’s planned first-light optical
spectrographs, G-CLEF and GMACS, will produce transformative early
science for users. However, the 25m GMT must provide access to its whole
20\,arcmin field of view if it is to fully realise its potential
$A\Omega$ advantage relative to the 30m TMT and the 39m ELT (see
Figure~1). With MANIFEST, GMT can offer 3-4$\times$ performance gains
for wide-field and survey spectroscopy.

\begin{figure}[b]
% \vspace*{-2.0 cm}
\begin{center}
 \includegraphics[width=0.6\textwidth]{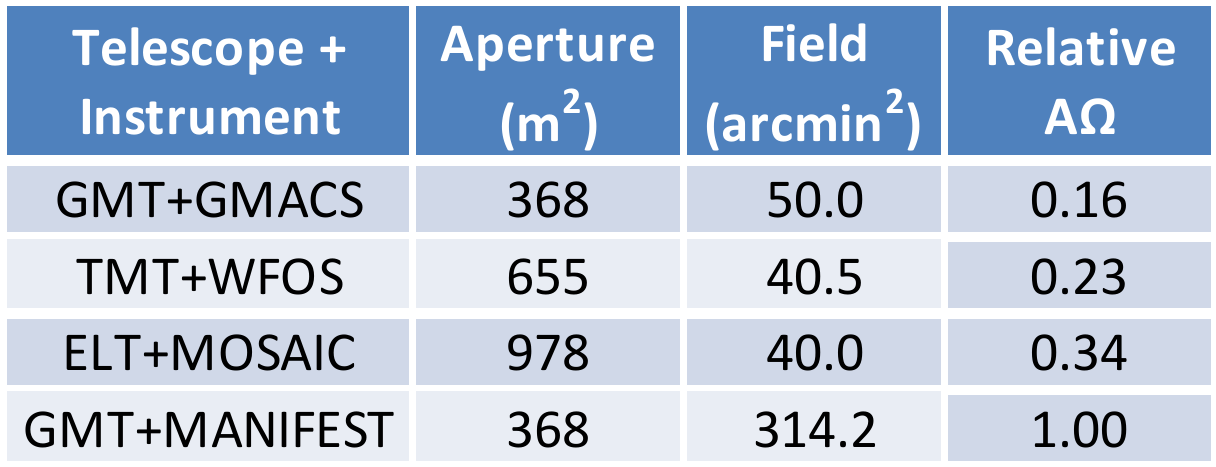} 
% \vspace*{-1.0 cm}
 \caption{Comparison of the relative $A\Omega$ (the product of telescope
   aperture and instrument field of view) for various ELT+spectrograph
   combinations, showing the factor of 3--4 gain offered by
   GMT+MANIFEST.} \label{fig1}
\end{center}
\end{figure}

GMT's wide field corrector and atmospheric dispersion compensator
(WFC/ADC) combined with the MANIFEST fibre system provides access to
GMT’s {\it entire} field of view for {\it multiple}
spectrographs---initially the GMACS optical medium-resolution
spectrograph and the G-CLEF optical high-resolution spectrograph, and in
future a near-infrared medium-resolution spectrograph such as the
proposed NIRMOS instrument. MANIFEST also provides a {\it versatile} set
of fibre feeds, ranging from single fibres through small image-slicers
to multiple integral field units. Detailed technical descriptions of
MANIFEST are provided by \cite{Saunders+2010}, \cite{Goodwin+2012},
\cite{Lawrence+2014}, \cite{Jacoby+2016} and \cite{Lawrence+2016}.

\begin{figure}
% \vspace*{-2.0 cm}
\begin{center}
 \includegraphics[width=0.7\textwidth]{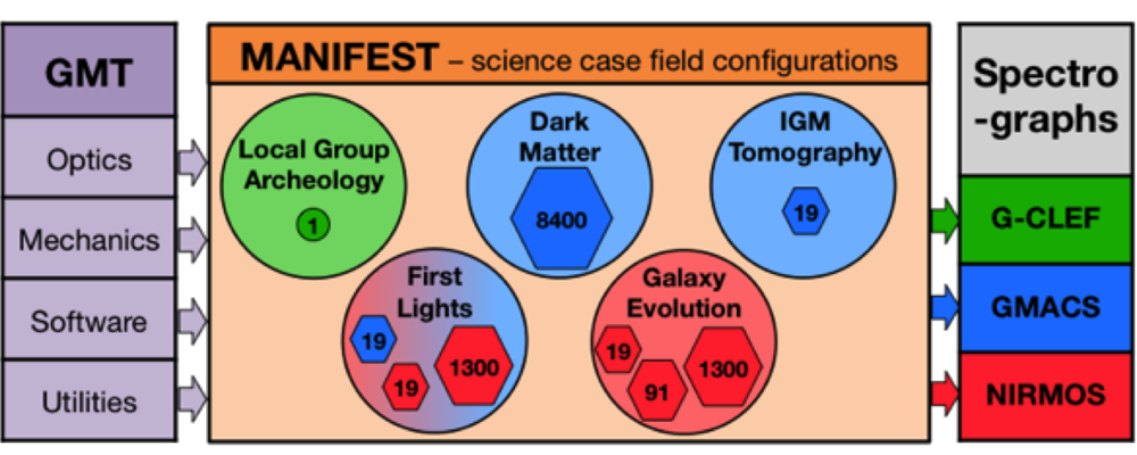} 
% \vspace*{-1.0 cm}
 \caption{Schematic showing MANIFEST's role in linking the GMT to the
   available spectrographs and the applicability of specific
   combinations of instruments and MANIFEST fibres/IFUs to various key
   science cases.}
   \label{fig2}
\end{center}
\end{figure}

Here I explore some of the early science that can be done by combining
GMT with MANIFEST and either GMACS or G-CLEF. As an overview,
Figure~\ref{fig2} provides a schematic showing some of the key science
cases for MANIFEST and their mapping to specific spectrographs via
various potential fibre or integral field unit (IFU) configurations. The
following paragraphs briefly summarize some of the key early science
that will be carried out with MANIFEST on GMT.

\begin{figure}[b]
% \vspace*{-2.0 cm}
\begin{center}
 \includegraphics[width=0.55\textwidth]{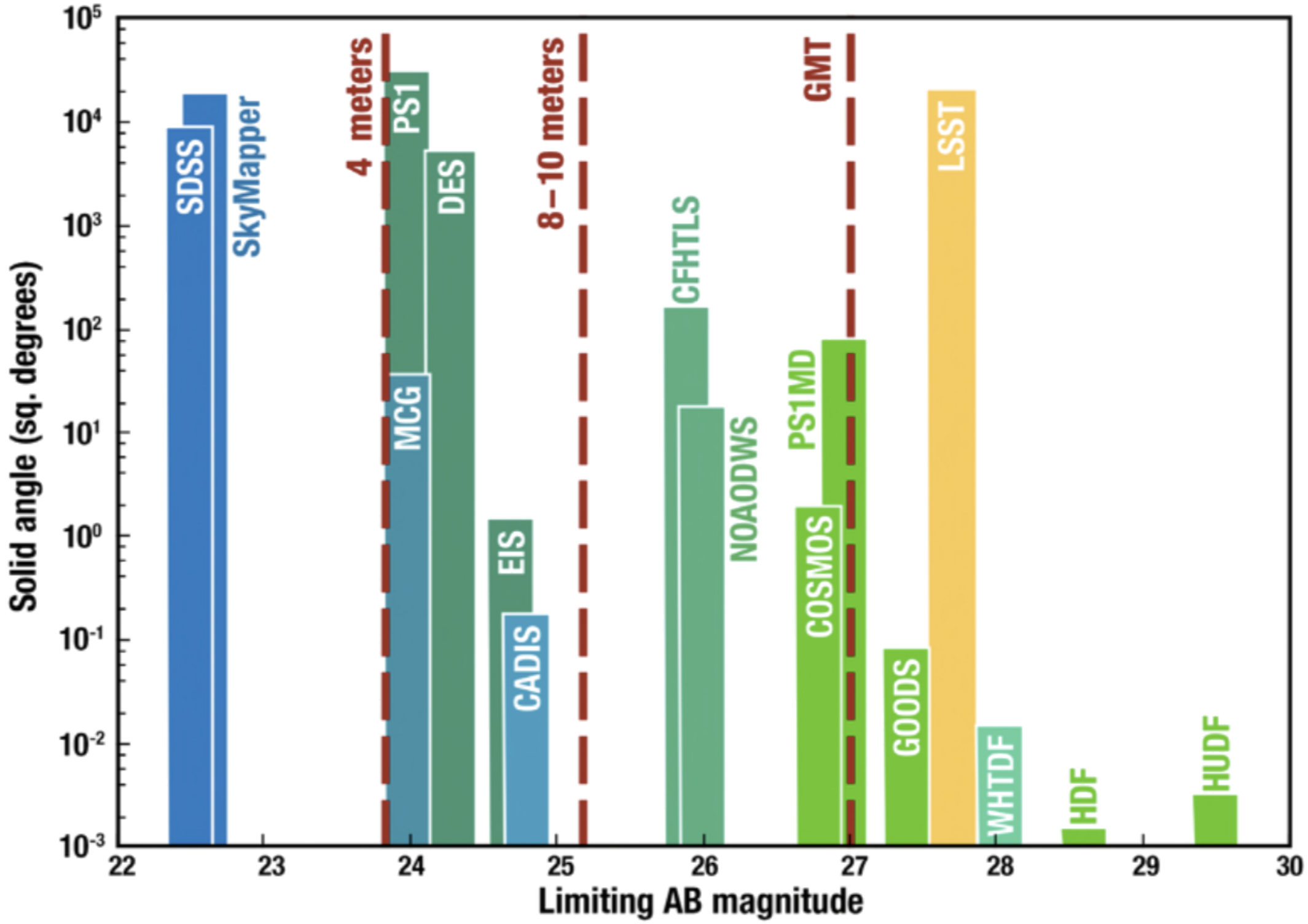} 
% \vspace*{-1.0 cm}
 \caption{Survey area and sensitivity ($5\sigma$ depth) of imaging
   surveys expected to be complete or underway at GMT first light,
   together with the corresponding spectroscopic limits ($5\sigma$
   continuum $I$-band detection) in 20 hours on a 4m, 8-10m and GMT.
   [\cite{Bernstein+2018}, Fig.1-10]}
   \label{fig3}
\end{center}
\end{figure}

{\bf Synergies with wide-field imaging surveys.}  GM with MANIFEST and
GMACS will reach the 5--10$\sigma$ detection limits of wide-field
imaging surveys (including LSST survey fields) in practical exposure times,
as shown in Figure~\ref{fig3}. MANIFEST and GMACS can thus measure
dynamics and abundances for almost any source in the LSST catalogue. The
observing efficiency offered by MANIFEST's wide field and high multiplex
means GMT will excel at exploiting imaging surveys.

{\bf Spectroscopic surveys at z\,$\boldsymbol{\sim}$\,0.5.} MANIFEST is
well-suited to spatially-resolved galaxy surveys at intermediate
redshifts. Currently, the SAMI multi-IFU system on the AAT 4m is
surveying 10$^3${--}10$^4$ galaxies at $z=0.05${--}0.1. In future,
MANIFEST and GMACS could study evolution in the resolved properties of
galaxies out to $z$\,$\sim$\,0.5 for similarly-sized samples. A
densely-sampled, volume-limited, spatially-resolved galaxy survey can
study detailed evolution over the last 5\,Gyr of (i)~mass and angular
momentum growth; (ii)~black hole feeding and feedback; (iii)~stellar
winds and outflows; and (iv)~kinematic morphology transformations.

\begin{figure}[t]
% \vspace*{-2.0 cm}
\begin{center}
 \includegraphics[width=0.75\textwidth]{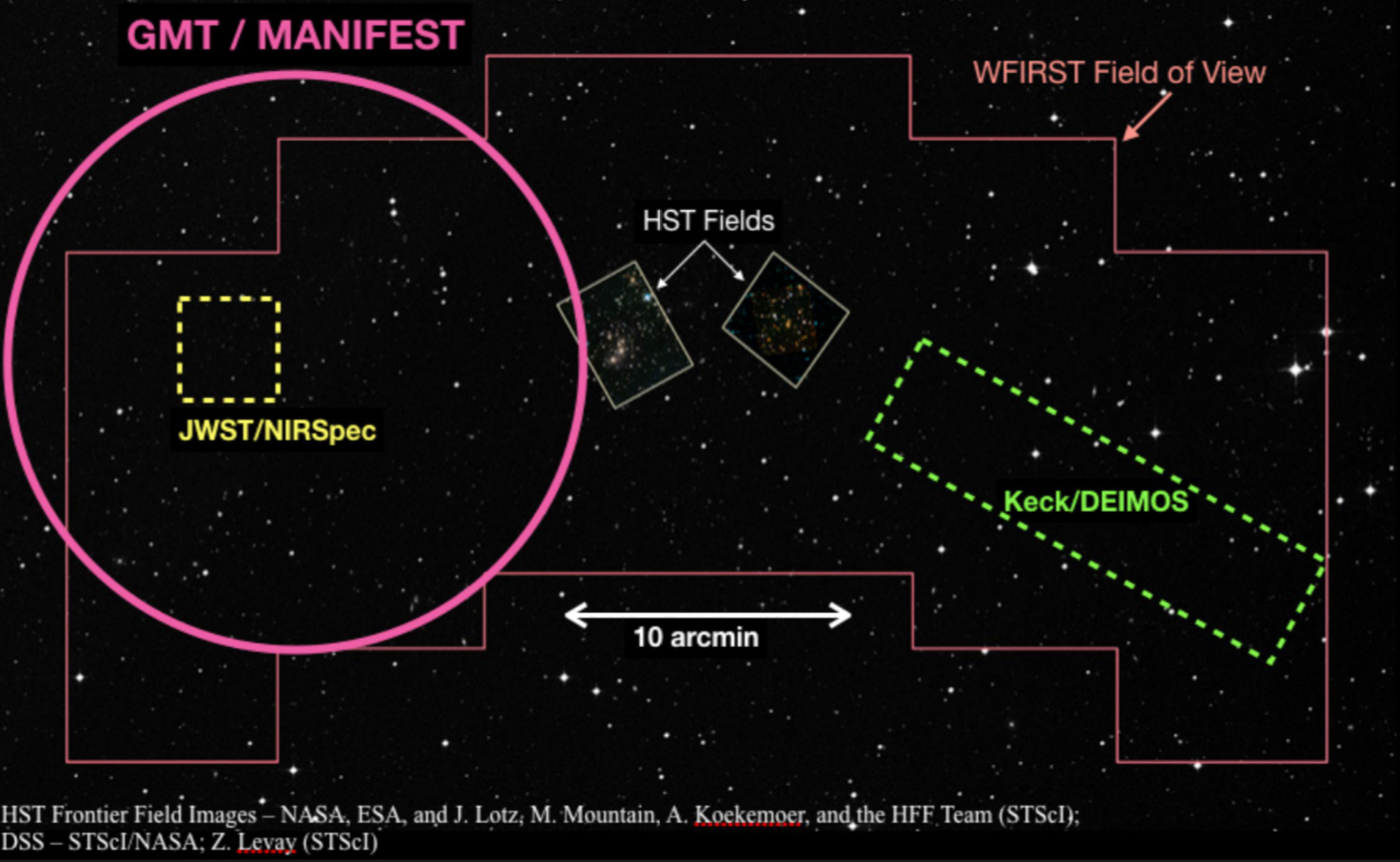} 
% \vspace*{-1.0 cm}
 \caption{The DSS image of the region around Abell 2744, inset with deep
   images from the HST Frontier Fields. The overlays show that
   MANIFEST's 20~arcmin field of view can cover almost all of a WFIRST
   field in just 3 shots.  [\cite{Bernstein+2018}, Fig.9-2]}
   \label{fig4}
\end{center}
\end{figure}

{\bf Spectroscopic surveys at z\,$\boldsymbol{>}$\,6.} GMT's large
aperture and wide field are well-suited to studying faint, high-redshift
galaxies. MANIFEST and GMACS provide an efficient way to follow up
$z$\,$>$\,6 galaxies with large surveys. A single deep
($H_{AB}=27$\,mag) WFIRST pointing will include of order 1000
$z$\,$>$\,6 galaxies---see \cite{Bernstein+2018}. MANIFEST and GMACS can
detect Ly\,$\alpha$ in all $z$\,$\sim$\,6--7 galaxies with line fluxes
$>$\,3\,$\times$\,10$^{-17}$\,erg\,s$^{-1}$\,cm$^{-2}$ with
$S/N$$>$5 in a 10-hour integration. As Figure~\ref{fig4} shows,
MANIFEST's 20~arcmin field of view can cover almost all of a WFIRST
field in just 3 shots.

{\bf IGM/CGM absorption-line surveys.} MANIFEST and GMACS will be able
to perform tomography on the intergalactic medium (IGM) and
circumgalactic medium (CGM) using Lyman-break galaxies. The aim is to
reconstruct the 3D small-scale structure of the IGM and CGM at high
redshifts by very densely sampling the Ly\,$\alpha$ forest over large
areas of sky. GMT can see the Ly\,$\alpha$ forest in the spectra from
faint (and therefore dense) Lyman-break galaxy samples at
2\,$<$\,$z$\,$<$\,3.5 (0.36--0.56\,$\mu$m). Using MANIFEST will double
the object multiplex and spectral resolution relative to GMACS alone,
and provide sub-Mpc spatial sampling of the IGM along each line of
sight.

{\bf Circumgalactic medium in emission.} With a large-field optical IFU,
MANIFEST and GMACS will enable studies of the CGM in emission around not
only QSOs but also inactive galaxies, where the ionizing flux is much
lower than in the vicinity of QSOs. Cosmological surface brightness
dimming means the sensitivity needed to detect faint Ly\,$\alpha$ flows
varies strongly with redshift; with a blue-sensitive CCD, MANIFEST and
GMACS will be able to target Ly\,$\alpha$ emission in and around
galaxies at $z$\,$\sim$\,2. The virial radius of a typical star-forming
galaxy at $z$\,$\sim$\,2 is about 90\,kpc, corresponding to an angular
extent of 11\,arcsec, so a large optical IFU is ideal for these studies.

{\bf Stellar chemistry and Galactic archaeology.}  Using MANIFEST in
combination with G-CLEF allows the study of stellar chemical abundances
in Local Group galaxies, and these abundances can be used to trace the
formation history of the Milky Way itself and nearby Local Group
galaxies, including the LMC, SMC, and about 40 dwarf galaxies within
about 1\,Mpc. A survey of 2000 stellar spectra at $S/N\sim30$ sampling
100 red giant branch stars in each of 20 Local Group galaxies would take
about 30 nights on GMT with MANIFEST and G-CLEF.  This combination can
also study the Galactic archaeology of the outer Milky Way disk via
chemical tagging and enable a faint extension of the current GALAH
survey on the AAT 4m, which is using chemical abundances for a million
bright stars to chemically tag coeval stellar associations in the inner
disk and bulge. With GMT, it should be possible to extend such surveys
to the outer disk and even the inner halo.

\vspace*{6pt}

\noindent In summary, by allowing GMT to exploit its full 20~arcmin
field of view, MANIFEST will make GMT the `wide-field ELT', with a
3--4$\times$ $A\Omega$ advantage over other ELTs. It will also provide
versatile high-multiplex/multi-IFU feeds for the GMACS and G-CLEF
first-light spectrographs. For galaxy evolution, MANIFEST and GMACS can
combine a survey of galaxies at the peak star formation epoch with
studies of gas flows between galaxies and the circumgalactic medium,
mapping both the emission from hot gas using integral field spectroscopy
and the absorption from cold gas with multi-object spectroscopy of
background sources. For galactic archaeology and near-field cosmology
using Local Group galaxies, MANIFEST and G-CLEF can obtain up to 40
simultaneous high-resolution optical spectra over a wide field, and so
produce detailed kinematic and chemical maps of the stellar populations
out to large radii in galaxies covering a broad range of masses and
morphologies. Consequently, MANIFEST should be a key element of the
early science programs carried out with GMT.

\begin{discussion}

\discuss{R.\ Egeland}{We heard about systems with 150 and 300 fibres
  ('starbugs'). What is setting the limit for the number of starbugs one
can implement?}

\discuss{M.\ Colless}{The systems with 150 to 300 starbugs are
  single-fibre prototypes. MANIFEST will probably have hundreds of IFU
  starbugs and over 1000 fibres. }

\discuss{B.\ Weiner}{Is there a tension between the fibre diameters
needed to do integrated high-throughput spectroscopy of galaxies/stars,
and that need for IFU spatially-resolved spectra? Are different
fore-optics needed?}

\discuss{M.\ Colless}{MANIFEST will have image-slicers for
  high-throughput integrated spectroscopy and IFUs for
  spatially-resolved spectroscopy. We expect to provide a variety of
  starbugs offering different feeds for different applcations and
  different instruments.}

\discuss{G.\ Bono}{Is the limiting magnitude you mentioned for MANIFEST
  for seeing-limited or AO-assisted observations?}

\discuss{M.\ Colless}{The limits mentioned apply to seeing-limited
  observations.}

\end{discussion}
    
\end{document}